\documentclass{cimento}

\newcommand{\Dbar}{\ensuremath{\overline{D}}}
\newcommand{\Dzb}{\ensuremath{\overline{D}^0}}
\newcommand{\Dz}{\ensuremath{D^0}}
\newcommand{\Dp}{\ensuremath{D^+}}
\newcommand{\Dm}{\ensuremath{D^-}}
\newcommand{\Dsp}{\ensuremath{D_s^+}}
\newcommand{\Dsm}{\ensuremath{D_s^-}}

\usepackage{graphicx}  
\title{Heavy Flavor Physics at CLEO-c}
\author{P.~U.~E.~Onyisi\from{ins:uc},
for the CLEO Collaboration}
\instlist{\inst{ins:uc} Enrico Fermi Institute, University of Chicago, Chicago, IL 60637, USA}
\PACSes{
\PACSit{12.15.Hh}{Determination of Kobayashi-Maskawa matrix elements}
\PACSit{13.20.Fc}{Leptonic, semileptonic, and radiative decays of charmed mesons}
\PACSit{13.25.Ft}{Hadronic decays of charmed mesons}}
\begin{document}

\maketitle

\begin{abstract}
The CLEO-c experiment, running at charm threshold, has measured many charmed meson properties.  Here I summarize results on leptonic and semileptonic decays of $D$ mesons, as well as measurements hadronic decay strong phases that are relevant to the extraction of the CKM angle $\gamma$ from $B$ decays.
\end{abstract}

\section{Introduction}
Studies of the interactions of matter with the electroweak symmetry breaking sector of the Standard Model have been very fruitful.  The interactions of quark mass eigenstates with the weak force --- a structure inherited from the Yukawa couplings of the quarks with the Higgs field --- must satisfy specific relationships in the Standard Model, and violations of those would signal physics beyond the SM.  The non-observation of such effects to date places stringent limits on the form of such scenarios.

All quarks except the top quark are only observed when confined inside hadrons, which is a regime where QCD is non-perturbative.  Relating observable hadron decays to ``short-distance'' weak dynamics requires precision understanding of the strong force.  Lattice QCD offers the prospect of a systematically-improvable method of calculating hadronic properties from first principles.  In the past decade theoretical and technological improvements (in particular the handling of virtual quark-antiquark pairs, the so-called ``unquenched'' calculations) have allowed the lattice to deliver predictions that are in many cases very precise, have no tunable parameters, and reliably estimate systematic uncertainties.  Before application of these results to extract electroweak parameters in the $B$ system, it is desirable to test them elsewhere, for example in charm.

The CLEO-c experiment at the CESR-c electron-positron collider collected large data samples in the charm threshold energy region.  Coupled with a well-understood detector, these samples enable tests of lattice predictions for charm hadron decays, as well as studies of many other topics.  Here I will discuss measurements of the meson decay constants $f_D$ and $f_{D_s}$ and branching fractions and form factors for the semileptonic decays $D^{0,+} \to (K,\pi) e^+ \nu_e$.  In addition, I will discuss studies of $D_s$ semileptonic decays, as well as studies of strong force-induced decay phases that are relevant for interferometry in the $B$ system.

\section{Detector and Data Samples}
The CLEO-c detector was a symmetric general purpose detector at the CESR-c $e^+ e^-$ collider.  The experiment is described in detail elsewhere \cite{cleo}. 
The relevant datasets for the following analyses were collected at center of mass energies of approximately 3.77 GeV (the peak of the $\psi(3770)$ resonance) and 4.17 GeV.  The former dataset is  used for \Dz\ and \Dp\ analyses, and the latter for $D_s$ physics.

At 3.77 GeV the only allowed open charm final states are $\Dz\Dzb$ and $\Dp\Dm$; at 4.17 GeV the only allowed states involving a $D_s$ meson are $D_s^+ D_s^-$ and $D_s^\pm D_s^{*\mp}$.  This enables the powerful tagging technique pioneered by Mark III \cite{markiii} which uses the presence of a fully reconstructed $D$ meson which decays to a tag mode to indicate the existence of its antiparticle.  This is the basis of the technique for measuring absolute branching fractions used in the analyses discussed below:
\begin{equation}
\mathcal{B}(D \to X) = \frac{N(\Dbar \to \textrm{tag}, D \to X)}{N(\Dbar \to \textrm{tag})}
\frac{\epsilon(\Dbar \to \textrm{tag})}{\epsilon(\Dbar \to \textrm{tag}, D \to X)}
\end{equation}
where the $\epsilon$ are the respective efficiencies.
There are other benefits to tagging: full reconstruction of the visible particles of an event allows a neutrino to be inferred; the removal of the particles constituting a tag strongly reduces the combinatorics (and hence backgrounds) of the rest of the event; and a judicious choice of \Dz\ tags allows the exploitation of quantum correlations of the initial state to measure phases.

\section{Leptonic Decays and Decay Constants}
The decays $X^+ \to \ell^+ \nu$ involve a hadronic current (which can be parametrized by the single scalar ``decay constant'' $f_X$) and a leptonic current, which is well understood in the electroweak model.  Consequently the branching fraction for such a decay can be written
\begin{equation}\label{form:lepbf}
 \mathcal{B}(X^+ \to \ell^+ \nu) = f_X^2 |V|^2 \frac{G_F^2}{8\pi} m_X m_\ell^2 \left(1-\frac{m_\ell^2}{m_X^2}\right)^2
\end{equation}
where $V$ is the relevant element of the CKM matrix connecting the valence quarks of $X$ (for \Dp\ and \Dsp\ this is $V_{cd}$ and $V_{cs}$, respectively).  Experiment can measure the quantity $f_X^2 |V|^2$; knowing the decay constant, we can obtain the CKM element, and vice versa.

In a na\"ive quantum mechanical picture, the decay constant can be thought of as a measure of the wave function of the meson at zero separation between the quarks.  This means it is relevant for processes where the relevant length scales are much smaller than the hadron size, such as the loop diagrams for $B_d^0$ and $B_s^0$ mixing.  The mass difference between $B_{(s)}$ eigenstates is proportional to $f_{B_{(s)}}^2$; as this is our primary source of information on $V_{td}$, reducing theoretical uncertainty is critical.


The measurements are discussed below.

\subsection{$\Dp \to \mu^+ \nu$}
\begin{figure}
\begin{center}
\includegraphics[width=.45\linewidth]{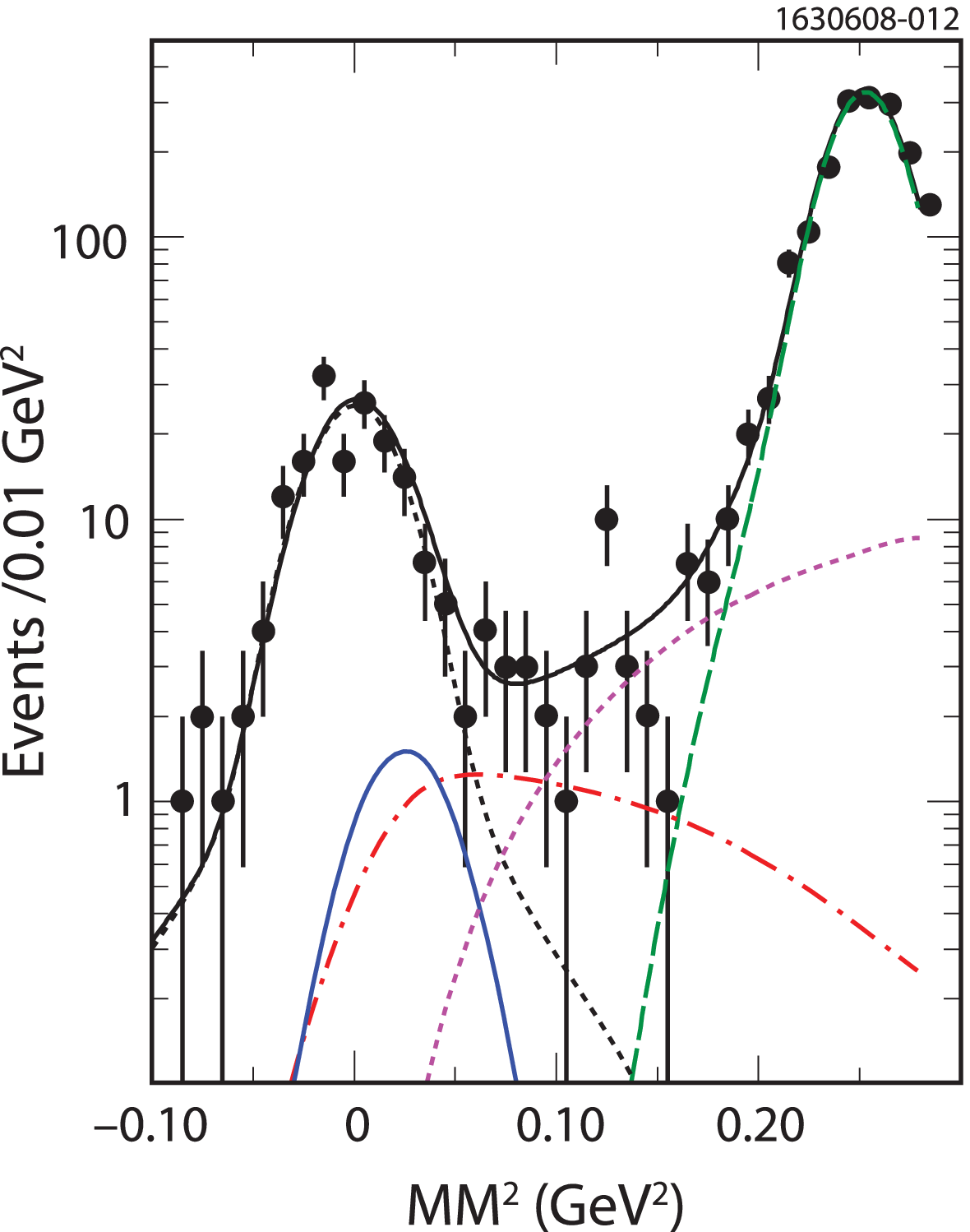}\hfill
\includegraphics[width=.49\linewidth]{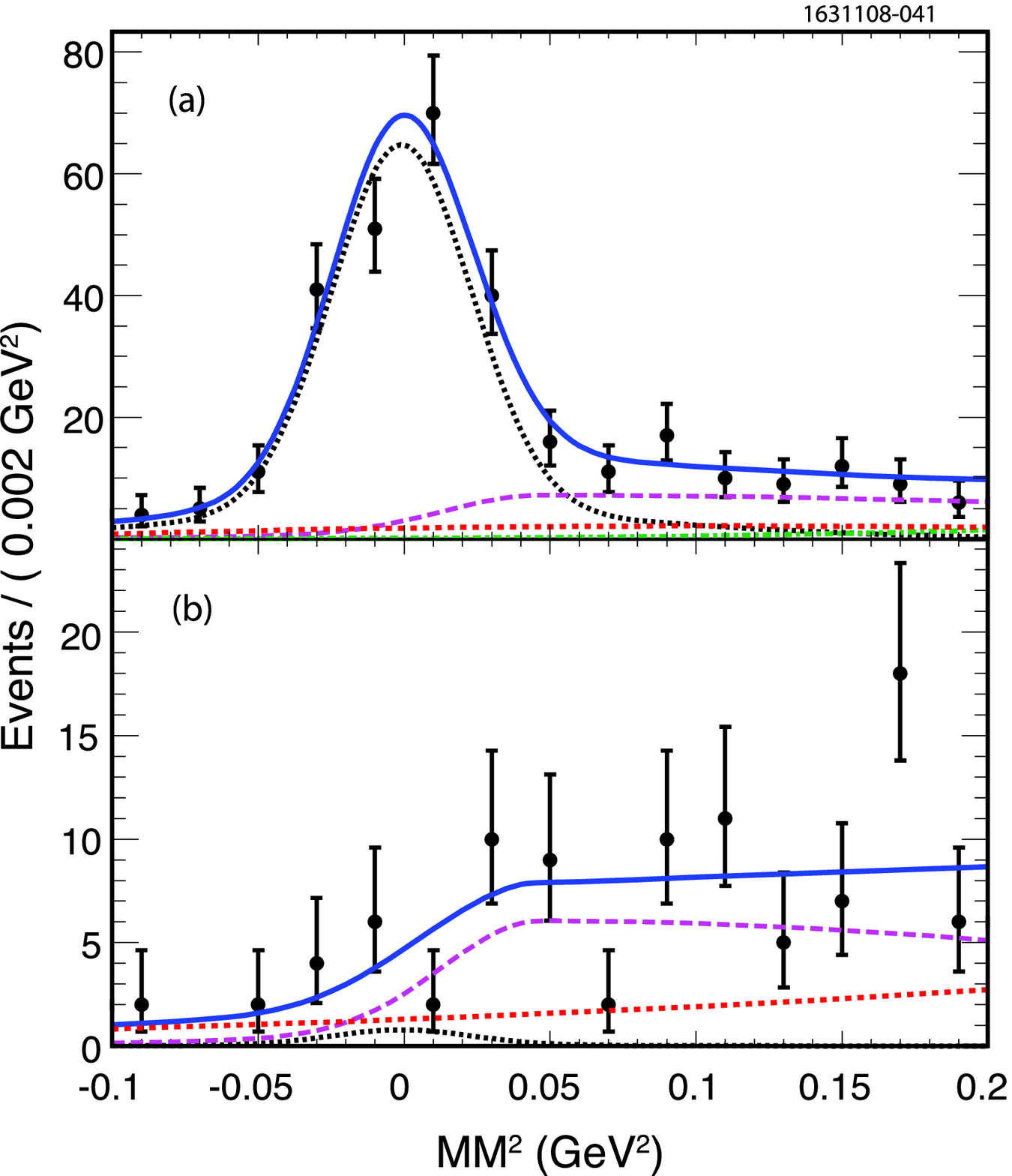}
\caption{\label{fig:dmumm} {\it Left:} distribution of $MM^2$ for $\Dp\to\mu^+\nu$ candidate events. The black solid line is a fit to the spectrum summing the following components: $\Dp\to\mu^+\nu$ signal (dotted black line peaking at zero); $\Dp \to \pi^+ \pi^0$ (solid blue); $\Dp \to \tau^+\nu$ (dot-dashed red); $\Dp \to \overline{K}^0 \pi^+$ (dashed green); and other three-body decays (dotted purple line rising towards the right of the plot). {\it Right:} Distribution of $MM^2$ for (a) $\Dsp\to\mu^+\nu$ and (b) $\Dsp \to \tau^+ \nu \to \pi^+ \bar \nu \nu$ candidate events. The blue solid line is a fit to the spectrum summing the following components: $\Dsp\to\mu^+\nu$ signal (dotted black line peaking at zero); $\Dsp \to \tau^+ \nu$ (long-dashed purple); other $\Dsp$ decays (dot-dashed green); and non-\Dsp\ decays (dashed red).}
\end{center}
\end{figure}

This analysis \cite{fd} uses the full 818 pb$^{-1}$ of 3.77 GeV data.  One of six hadronic \Dm\ decays is reconstructed as a tag\footnote{Charge conjugate reactions are implied.} which sets the initial number of \Dm\ decays.
Exactly one track is allowed aside from those composing the track; this is taken as the muon candidate, and must have deposited less than 300 MeV in the electromagnetic calorimeter and not have been considered a viable kaon candidate by the particle identification algorithm.  If the event has extra calorimeter energy, it is vetoed, to eliminate one prong hadronic \Dp\ decays with $\pi^0$s.  The four-vectors of the initial state, tag, and muon candidate are then combined to form the missing mass squared $MM^2 \equiv (p_0 - p_{\Dm} - p_{\mu^+})^2$.
This peaks at zero (the neutrino mass squared) for signal events, as shown in fig.~\ref{fig:dmumm}.  A fit is performed to the spectrum including the signal and various background components.  Some $\Dp \to \tau^+ \nu$ events are expected to leak into this plot, but the number is too low to fit for explicitly and that component is fixed relative to the $\Dp \to \mu^+ \nu$ contribution according to the SM expectation for the ratio.

\subsection{$\Dsp \to \mu^+ \nu$, $\tau^+ \nu$ $(\tau^+ \to \pi^+ \bar\nu)$}
This measurement \cite{fds1} proceeds similarly to the $\Dp \to \mu^+ \nu$ analysis.  The full 600 pb$^{-1}$ dataset at 4.17 GeV is used.  At this energy the dominant \Dsp\ production mode is $e^+ e^- \to D_s^\pm D_s^{*\mp}$; the $D_s^{*\mp}$ then decays to $D_s^\mp\gamma$ (94.2\%) or $D_s^\mp \pi^0$ (5.8\%) \cite{pdg08}.  Thus, compared to the $\Dp$ case, there is an extra particle that must be considered when forming the missing mass squared.  Only the photon transition is considered in this analysis.

Nine hadronic \Dsm\ tag modes are used to obtain the parent sample of events.  One extra track is allowed, as well as a transition photon candidate; additional calorimeter energy is vetoed.  The extra track is required to be either muon-like (less than 300 MeV deposited in the calorimeter) or pion-like (more than 300 MeV deposited, but track is not an electron candidate).  A kinematic fit is performed which uses multiple constraints to improve the $MM^2$ resolution.  

The distribution of the $MM^2$ is shown in fig.~\ref{fig:dmumm}.

\subsection{$\Dsp \to \tau^+ \nu$ $(\tau^+ \to e^+ \nu\bar\nu)$}
\begin{figure}
\begin{center}
\includegraphics[width=.5\linewidth]{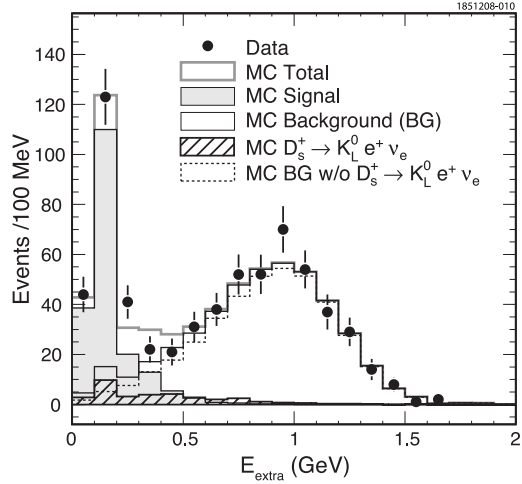}
\caption{\label{fig:dsptau} Distribution of extra calorimeter energy for $\Dsp \to \tau^+ \nu \to e^+ \bar \nu \nu \nu$ candidate events.  The signal region is 0-400 MeV.  The signal shape is the lightly shaded histogram peaking just above zero.  The total expected background is the open histogram.  The primary background to the measurement, $\Dsp \to K^0_L e^+ \nu$, is the hatched histogram.}
\end{center}
\end{figure}

This measurement \cite{fds2} uses a different technique from the previously-discussed decay constant measurements.  In the \Dp\ case, the missing mass squared variable serves to separate the signal $\Dp \to \mu^+ \nu$ from, in particular, $K^0_L$ backgrounds.  For the \Dp\ the signal is Cabibbo-suppressed and the background (e.g.\ $\Dp \to K^0_L \pi^+$) is Cabibbo-favored.  In the \Dsp\ case this is largely reversed.  Reconstructing a \Dsp\ tag and an electron and imposing an additional track veto selects the signal decay as well as semileptonic decays with neutral hadrons, but critically most of these result in additional photons.  Requiring only small amounts of additional calorimeter energy strongly discriminates for the signal, as shown in fig.~\ref{fig:dsptau}; the main remaining background is the Cabibbo-suppressed $\Dsp \to K^0_L e^+ \nu$.

\subsection{Results and Combination}
The results of the leptonic branching fraction measurements and corresponding decay constants are shown in table~\ref{tbl:fdresults}.  The values of input parameters used to obtain these values are listed in the relevant papers \cite{fd,fds1,fds2}. 

\begin{table}
\caption{\label{tbl:fdresults} CLEO-c measurements of \Dp\ and \Dsp\ leptonic decay branching fractions and decay constants, compared to lattice QCD predictions from the HPQCD and UKQCD collaborations \cite{hpqcd}.}
\begin{tabular}{lcc}
\hline
& CLEO-c Result & Lattice QCD\\
\hline
$\mathcal{B}(\Dp \to \mu^+ \nu)$ & $(3.82 \pm 0.32 \pm 0.09) \times 10^{-4}$ \\
$\mathcal{B}(\Dsp \to \mu^+ \nu)$ & $(5.65 \pm 0.45 \pm 0.17) \times 10^{-3}$ \\
$\mathcal{B}(\Dsp \to \tau^+ \nu)$ (from $\tau^+ \to \pi^+ \bar\nu$) & $(6.42 \pm 0.81 \pm 0.18) \times 19^{-2}$\\
$\mathcal{B}(\Dsp \to \tau^+ \nu)$ (from $\tau^+ \to e^+ \nu \bar\nu$) & $(5.30 \pm 0.47 \pm 0.22) \times 10^{-2}$\\
$f_{\Dp}$ & $205.8 \pm 8.5 \pm 2.5$ MeV& $207 \pm 4$ MeV\\
$f_{\Dsp}$ (combined) & $259.5 \pm 6.6 \pm 3.1$ MeV & $241 \pm 3$ MeV\\
$f_{\Dsp}/f_{\Dp}$ & $1.26 \pm 0.06 \pm 0.02$ & $1.162 \pm 0.009$\\
\hline
\end{tabular}
\end{table}

\section{Exclusive Semileptonic Decays}
Exclusive semileptonic decays have a more involved parametrization than leptonic decays, as they involve at least three particles in the final state.  The partial width for the decay $X \to X' \ell \nu$, where $X$ and $X'$ are pseudoscalars, can be written as
\begin{equation}\label{eqn:semilep}
\frac{d\Gamma(X \to X'\ell \nu)}{dq^2} = \frac{G_F^2}{24\pi^3}\left[f_+^{X\to X'}(q^2)|V|\right]^2 p_{X'}^3
\end{equation}
in the limit where the charged lepton mass is negligible.  In eq.\ref{eqn:semilep},
 $q^2$ is the invariant mass squared of the $\ell \nu$ system, $|V|$ is the relevant CKM matrix element for the weak transition, and $f_+^{X \to X'}$ is a form factor encapsulating the hadronic physics.  As in the leptonic decay case, input for $|V|$ or $f_+$ allows determination of the other.

\subsection{\Dz\ and \Dp\ Decays}
\begin{figure}
\begin{center}
\includegraphics[width=.48\linewidth,height=5.5cm]{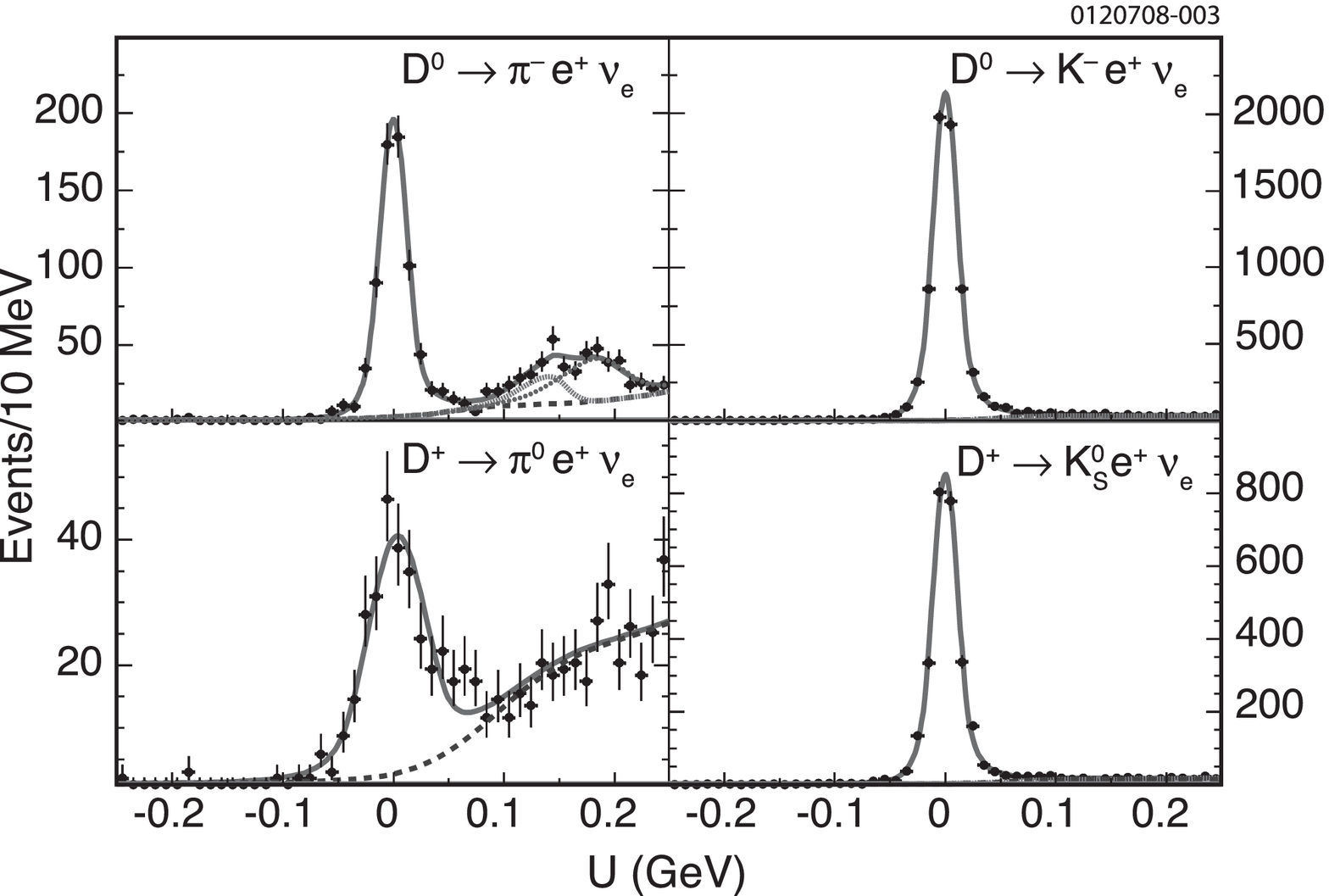}\hfill
\includegraphics[width=.48\linewidth,height=5.5cm]{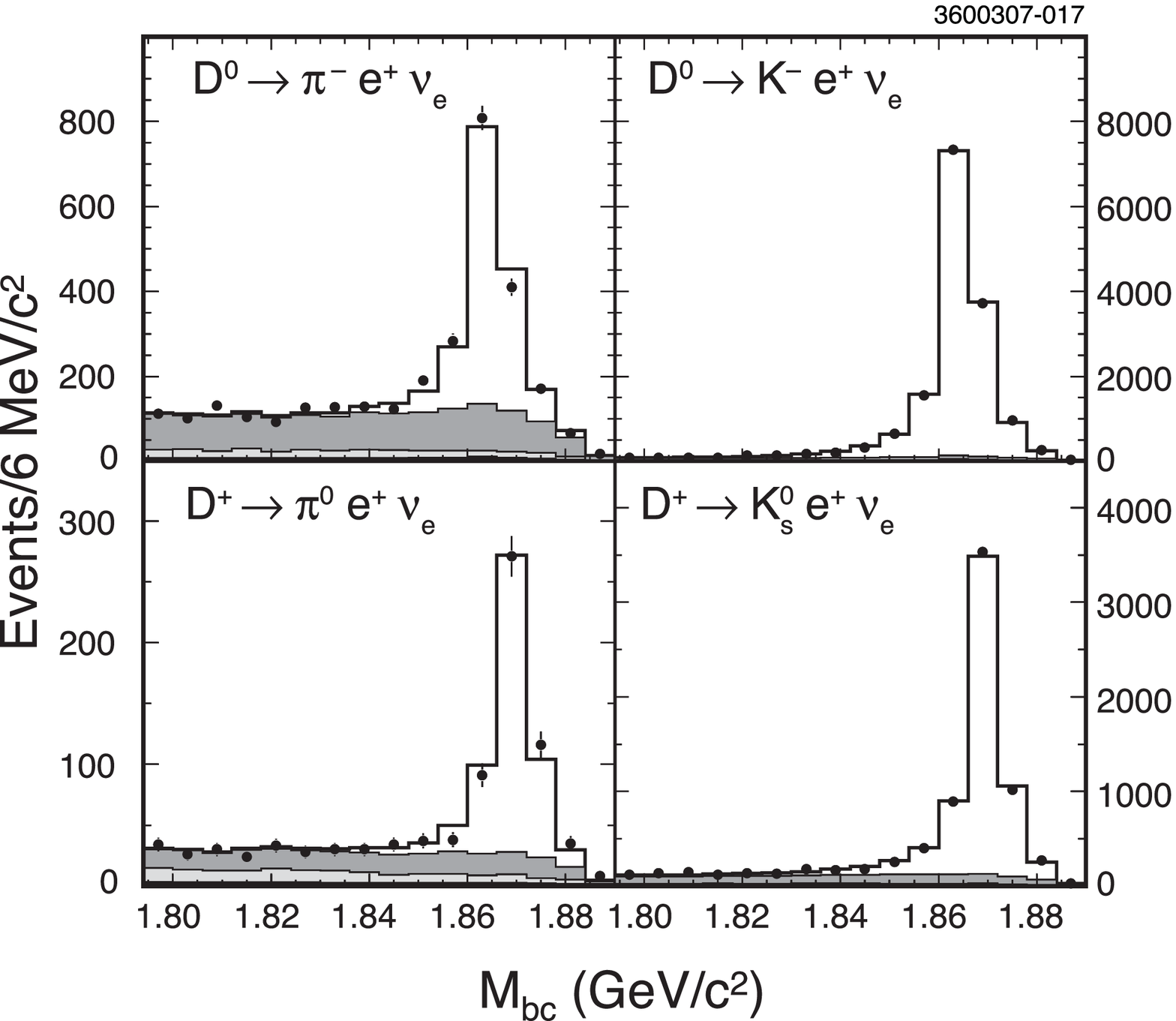}
\caption{\label{fig:semi} Signals for $D$ semileptonic decays in 281 pb$^{-1}$ of data. {\it Left:} distribution of $U\equiv E_{miss} - |\vec{p}_{miss}|$ in tagged analysis.  {\it Right:} distribution of $M_\mathrm{bc}$ in neutrino reconstruction analysis.}
\end{center}
\end{figure}

Two separate studies of $D \to (K,\pi) e^+ \nu$ were performed on 281 pb$^{-1}$ of 3.77 GeV data.  The ``tagged'' analysis \cite{dtagged} reconstructs hadronic decays of one $D$ in the event to establish the base sample.  A hadron ($K^\pm$, $\pi^\pm$, $K_S^0$, $\pi^0$) and an electron candidate are then selected, and the missing energy $E_{miss}$ and momentum $\vec{p}_{miss}$ are determined.  From these the variable $U \equiv E_{miss} - |\vec{p}_{miss}|$ is computed, which for correctly reconstructed events with neutrinos is approximately zero.  
The ``neutrino reconstruction'' analysis \cite{duntagged}
uses the near-hermeticity of the detector to attempt to reconstruct all visible particles in an event; if the missing four momentum is consistent with the neutrino mass, it is considered a neutrino candidate, and an attempt is made to combine it with electron and kaon or pion candidates to make a $D \to (K,\pi) e \nu$ candidate.  In this case the $D$ candidates are discriminated from background by looking at the variables $\Delta E \equiv E_D - E_\mathrm{beam}$ and $M_\mathrm{bc} \equiv \sqrt{E_\mathrm{beam}^2 - |\vec{p}_D|^2}$.  Fig.~\ref{fig:semi} shows the signals for both analyses.  

The yields as a function of $q^2$ are used to derive measurements of the form factors.  These are fit to several parametrizations: the simple pole model assuming $D_s^*/D^*$ dominance, the ``modified pole'' model \cite{bk}, and a series expansion \cite{becherhill}.  All fits describe the data reasonably as long as all parameters are allowed to float; however, for example, the implied $D_s^*/D^*$ pole masses in the pole models are many standard deviations from the physical values.  Reasonable agreement on the form factor shape and normalization is found with a lattice QCD prediction from the FNAL, MILC, and HPQCD collaborations \cite{fnal}.  Using lattice predictions for $|f_+(0)|$, values for $|V_{cd}|$ and $|V_{cs}|$ are also obtained, which are limited by lattice uncertainties.

\subsection{\Dsp\ Decays}
Using 310 pb$^{-1}$ of data, CLEO-c has made the first absolute measurement of the semileptonic decay branching fractions $\mathcal{B}(\Dsp \to X e^+ \nu)$ where $X \in (\phi, \eta, \eta', K^0, K^{*0}, f_0 \to \pi^+ \pi^-)$ \cite{dssemi}.  In contrast to previous measurements these are not ratios to a hadronic decay of the \Dsp.  In addition this is the first observation of the Cabibbo-suppressed modes $K^0 e^+ \nu$ and $K^{*0} e^+ \nu$.  The $\eta$ and $\eta'$ branching fractions provide useful information on the $\eta-\eta'$ mixing angle and glueball component \cite{etamix}.

\section{Strong Phases for $\gamma/\phi_3$ Measurements}
Of the three angles of the unitarity triangle, $\gamma$ has the largest uncertainties on its direct measurement.  A clean measurement of $\gamma$ from tree decays can be made by exploiting interference between the decays $b\to c \bar u s$ and $b \to u \bar c s$.  These correspond to decays $\overline{B} \to D K$ and $\overline{B} \to \overline{D} K$; since $D^0$ and $\overline{D}^0$ can decay to common final states, the interference can be observed.  Such final states include $K^- \pi^+$ (interference between Cabibbo-favored and doubly-Cabibbo-suppressed decays) \cite{ads} and $K_S^0 \pi^+ \pi^-$ (Cabibbo-favored in both cases, but populating different parts of phase space) \cite{dalitz}.  The total observed interference depends on $D$ decay dynamics --- specifically phases between $D$ and $\overline{D}$ decays to the same final state that are induced by the strong force.  Because $B$-factories observe definite flavor in $D^0$ decays (as they tag the soft pion in $D^{*+} \to D^0 \pi^+$), they cannot directly observe these phases.  In the case of decays to common three-body final states, the phases can be estimated by using models for the resonant substructure of the decays, but this leaves a residual model uncertainty.

\begin{figure}
\begin{center}
\includegraphics[width=.8\linewidth]{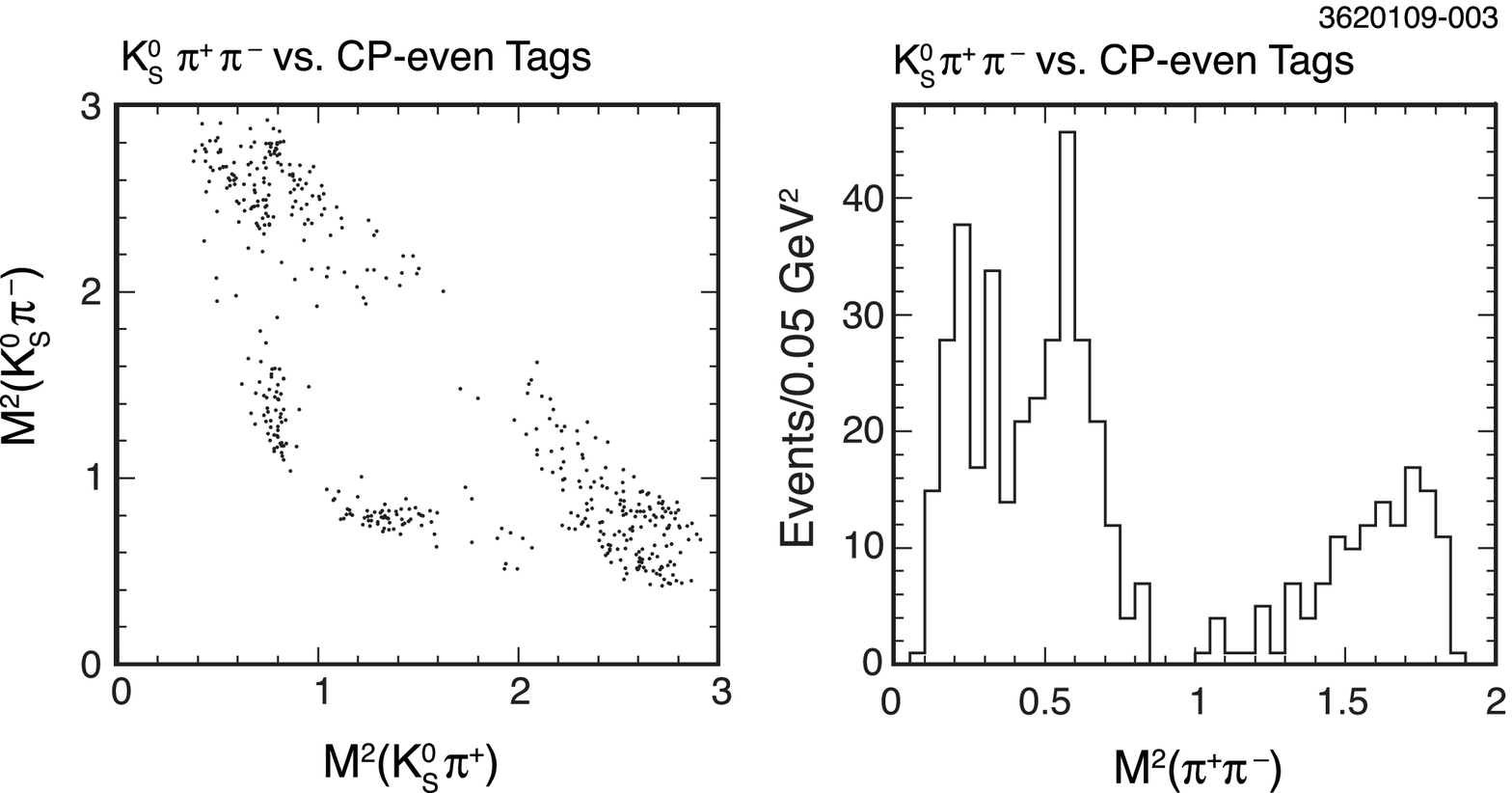}\\
\includegraphics[width=.8\linewidth]{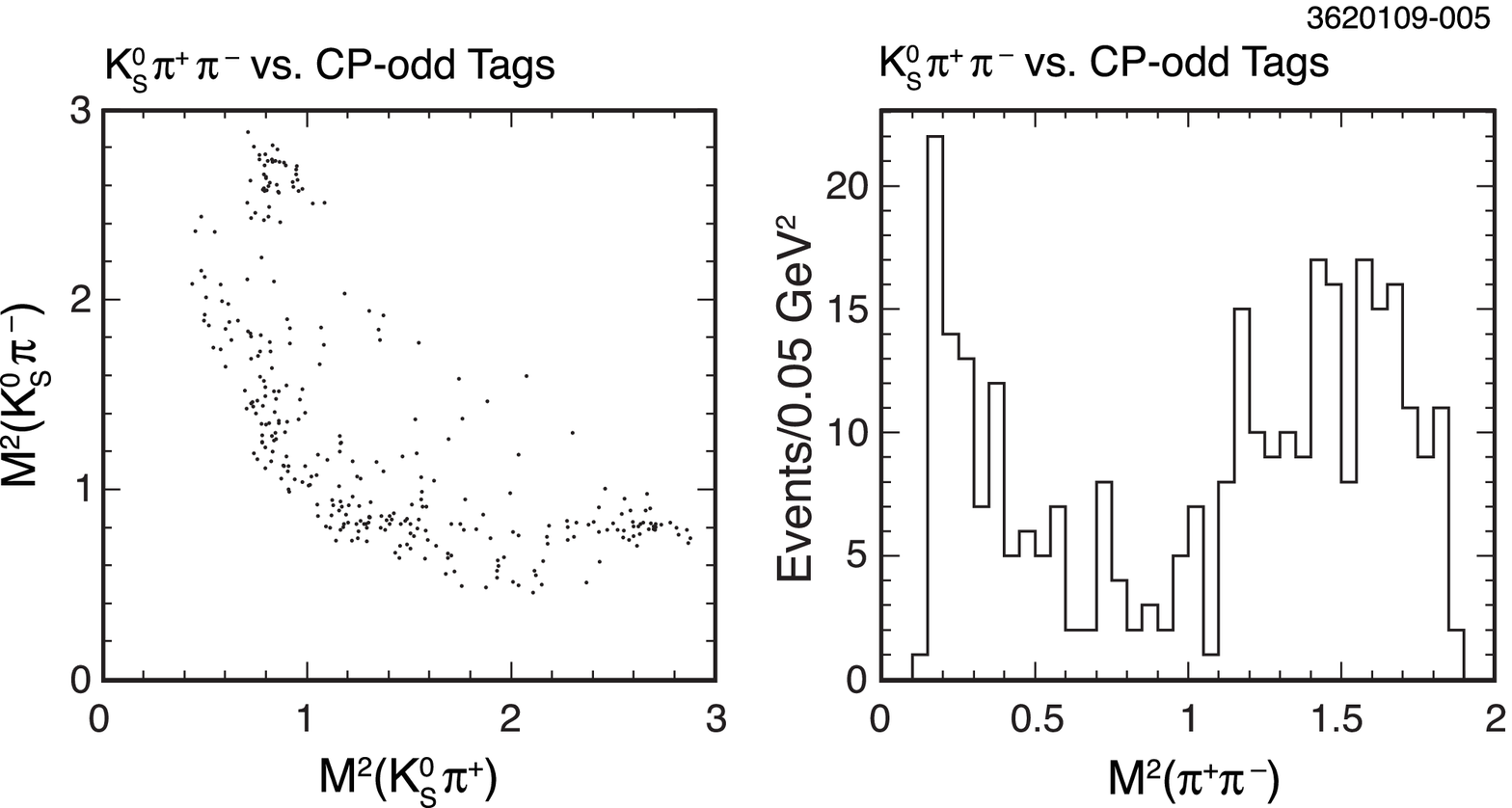}
\end{center}
\caption{\label{fig:cpcorr}Effects of $CP$ correlation on the Dalitz plot of the decay $\Dz \to K^0_{S} \pi^+\pi^-$.  The $K^0_S \rho$ component, clearly visible when $K^0_S \pi^+\pi^-$ recoils against a $CP$-even tag, disappears opposite a $CP$-odd tag.}
\end{figure}

Production of $\Dz\Dzb$ pairs at threshold provides unique access to the phase information.  The initial state is strongly constrained ($J^{PC} = 1^{--}$) and so the decays of the two $D$ mesons are correlated.  One obvious correlation is flavor-antiflavor (in the absence of mixing); a less obvious one is $CP$ correlation: if one $D$ decays to a $CP$ eigenstate (for example $K^- K^+$), the other must decay to a state of opposite $CP$.  This projects out a linear sum of the \Dz\ and \Dzb\ flavor eigenstates, which then interfere.  Comparing the rates and dynamics of the same decay when it happens opposite flavor and $CP$ eigenstates directly probes the strong phases in \Dz\ decays without model systematics.  The dramatic effect of these correlations is shown in fig.~\ref{fig:cpcorr}.

Here I briefly summarize CLEO-c results relevant to $\gamma$ measurements.  These results are statistically limited by the number of reconstructed $CP$ eigenstate decays.

\subsection{Strong Phase Between $\Dz\to K^- \pi^+$ and $\Dzb \to K^- \pi^+$}
The relative phase $\delta$ between the decays $\Dz\to K^- \pi^+$ and $\Dzb \to K^- \pi^+$ is relevant for the $\gamma$ measurement method of Ref.~\cite{ads}.  It also relates the \Dz\ mixing parameters $y$ and $y'$.  The differences in the effective branching fraction for $K^- \pi^+$ decay opposite $CP$-even and -odd eigenstates, semileptonic decays (which unambiguously determine the charge of the decaying charm quark), and $K^+ \pi^-$ are sensitive to $\delta$ and in principle to $\Dz$ mixing parameters as well.  With 281 pb$^{-1}$ of data, CLEO-c has made the first measurement of $\cos \delta$, finding it to be $1.03 ^{+0.31}_{-0.17} \pm 0.06$ \cite{tqca}.

\subsection{Effective Strong Phases In Multibody Decays}
The $\gamma$ measurement method using $K^- \pi^+$ decays can be extended to other decays with larger branching fractions \cite{atwoodsoni}.  In this case the relative phase depends on the decay kinematics, and the Cabibbo-favored and doubly-Cabibbo-suppressed decays will not have complete overlap over the phase space.  These effects can be subsumed in an effective average phase and a ``coherence factor'' which reflects the dilution of total interference relative to the expectation for a simple two-body decay.  CLEO-c has measured these for the $\Dz \to K^- \pi^+ \pi^0$ and $K^- \pi^+ \pi^+ \pi^-$ decays, observing significant coherence in the former \cite{coherencefactor}.

\subsection{Phase Space-Dependent Measurements}
One can proceed beyond the averaging approximation above and obtain relative \Dz--\Dzb\ phases as a function of decay kinematics by observing how $CP$ tagging affects Dalitz plots.  CLEO-c has performed this measurement for the $K_{S,L}^0 \pi^+ \pi^-$ decay \cite{kspipi} and work is underway for the $K_{S,L}^0 K^- K^+$ mode.  Fig.~\ref{fig:cpcorr} shows the effect of the $CP$ correlations on the $\Dz \to K_S^0 \pi^+\pi^-$ Dalitz plot.  Up to small effects $K_L^0 \pi^+ \pi^-$ is expected to have a $CP$ structure opposite that of $K_S^0 \pi^+ \pi^-$, and similarly for $K_L^0 K^- K^+$; CLEO-c is able to reconstruct $K^0_L$ candidates based on missing energy and momentum, so these decays can add to the measurement statistics for the phases although the $K_S^0$ modes are the ones relevant for $B$ factory measurements.

\subsection{Impact on $\gamma$ Measurement}
The impact of CLEO-c results on future analyses enabled by the large  dataset expected from LHCb has been studied.  The $\Dz \to K_S^0 \pi^+ \pi^-$ analysis is expected to reduce the current 7--9$^\circ$ model uncertainties from BaBar and Belle measurements \cite{bfacgamma} to around 2$^\circ$ \cite{kspipi,lhcb1}.  The $K^- \pi^+$ and multibody coherence factor measurements are projected to improve the 10 fb$^{-1}$ precision of LHCb in $B \to D K$ by 8--35\% (depending on unknown $B$ decay parameters) to 2.2--3.5$^\circ$ \cite{lhcb2}.




\acknowledgments
The author wishes to thank J.~Libby, C.~S.~Park, and P.~Naik for helpful discussions.  This work was partly supported by a Fermi Fellowship from the University of Chicago.

\end{document}